\documentclass[10pt]{article}
\usepackage[margin=.6in]{geometry} 
\usepackage{authblk}
\usepackage{graphicx}
\usepackage{amstext}
\usepackage{float} 

\begin{document}

\title{Semantic Navigation for AI-assisted Ideation}

\author{Thomas Sandholm, Sarah Dong, Sayandev Mukherjee, John Feland, Bernardo A. Huberman}
\affil{NextGen Systems, CableLabs, Santa Clara, CA}

\maketitle

\begin{abstract}
We present a novel AI-based ideation assistant and evaluate it
in a user study with a group of innovators.
The key contribution of our work is twofold: we propose
a method of idea exploration in a constrained domain by means of LLM-supported semantic navigation
of problem and solution spaces, and employ novel automated data input filtering
to improve generations. We found that semantic exploration is preferred to the traditional prompt-output
interactions, measured both in explicit survey rankings, and in terms of innovation assistant
engagement, where $2.1$x more generations were performed using semantic exploration. We also show that filtering input data with metrics such as 
relevancy, coherence and human alignment leads to improved
generations in the same metrics as well as enhanced quality of
experience among innovators. 
\end{abstract}

\section{Introduction}\label{sec:introduction}
The process of innovation in an organization usually begins with brainstorming, either individually or in a group.  Group brainstorming has been shown to suffer from productivity loss compared to working individually~\cite{diehl1987productivity} but also has benefit of engaging the organzational memory~\cite{sutton1996}.  One approach to boost overall innovation by enabling individual creativity while benefiting from organization (group) memory is to use an AI assistant as a tool allowing the individual to brainstorm by interacting with a Large Language Model (LLM)~\cite{vaswani2017} without social anxiety~\cite{hwang2021}. 

Note that given the nature of the creative process, the ideation input to the LLM tends to be noisy, and 
poor input gives poor output. A pre-trained general-purpose LLM asked to assist in ideation favors the most widely-circulated public-domain ideas, which hinders exploration and can lead to design fixation~\cite{wadinambiarachchi2024}. Adjusting the ``temperature'' parameter of the LLM allows for more surprising or creative output, but outside of a narrow range, yields more irrelevant and incoherent output.
Constraining innovation to a particular domain to get more relevant output presents the challenge of re-training or fine-tuning the LLM with potentially proprietary inputs, and repeating the process frequently, as innovations are constantly evolving and based on the latest findings and extensive organizational memory of past innovations. 

In this work, our main contributions are: (a) describing and providing a method to navigate and explore problem and solution spaces in a proprietary constrained domain during ideation assisted by an LLM (Section~\ref{sec:model}), 
(b) using efficient fine tuning of small models (Section~\ref{sec:motivation}) with automated filtering of high quality inputs to improve generations (Section~\ref{sec:cleaning}), and (c) a user study evaluating our ideation assistant with innovators (Section~\ref{sec:userstudy}). 

The semantic navigation in our model is achieved by allowing reverse solution-to-problem mappings in addition to the more typical
problem-to-solution mapping during ideation. We can traverse back to a problem statement from the generated solution as well as input pre-existing semantically related problem statements to generate additional solutions. 
This not only helps with exploration but also allows for refining problem statements, a fundamental stage of early ideation.

The use of smaller models that are feasible to host locally on affordable GPUs allows us to more (cost and compute) efficiently re-tune with new inputs while keeping proprietary data local.

Our automated input filtering method relies on measuring relevancy between prompt and output, as well as coherence of the generated text. Moreover, we apply concepts from the RLHF methodology~\cite{ouyang2022} to train a reward model that can be reapplied to filter new, unseen inputs.

We implemented and deployed our semantic navigation model in an ideation assistant accessed through a Slack bot (Section~\ref{implementation}).

The user study allowed us to evaluate the experience of innovators implicitly by monitoring their usage in logs as well as explicitly in surveys. This allowed users to evaluate the tool as an ideation assistant and integrate the benefits of group brainstorming (organizational memory) into the higher performing individual brainstorming process.  The surveys also
provided open-ended feedback that we could analyze qualitatively (Section~\ref{sec:conclusion}).

\section{Related Work}\label{sec:relatedwork}
LLM-assisted idea generation~\cite{di2022,chakrabarty2023,liu2023,wadinambiarachchi2024} is the latest in an extensive history of efforts in machine- and AI-assisted ideation~\cite{hwang2021,koch2019,shneiderman2002,yu2011}.

Early creativity support tools~\cite{shneiderman2002} targeted refining of existing innovations, 
exploration of related work to accelerate innovation, and connecting creators to enable collaborative ideation~\cite{yu2011}. A cooperative contextual bandit was proposed as an interactive design ideation tool in~\cite{koch2019}.
The impact that such tools have on the ideas produced has been investigated in a 
number of user studies, revealing, for example, that an AI-bot ideation
collaborator could improve both the quantity and quality
of generated ideas compared to human facilitators by
removing social pressure, anxiety, and bias~\cite{hwang2021};
and the LLM inference times while recursively exploring 
related research problems could pace the ideation
process for a better and more creative ideation 
experience~\cite{liu2023}. 

Using LLMs as a creative writing assistant was explored in
~\cite{chakrabarty2023}, where user studies showed that the
LLM was considered helpful in the writing translation 
(review) task, but more as an editor than an idea generator. A recent large-scale comparison of humans vs.~LLMs for research ideation~\cite{si2024} showed that LLM-generated research ideas written in a grant proposal format are judged (by humans) to be more novel (although less diverse and less feasible) than human expert ideas.
In~\cite{di2022} the authors propose a tool that leverages prompt 
engineering to expand, rewrite and combine innovations suggested by users,
demonstrating the multi- and general-purpose natural language capabilities of
LLM foundation models.  A train-of-thought methodology called Tree-of-Thought (ToT) was proposed for LLM-assisted problem solving in~\cite{long2023}. ToT was designed as a fully automatic process to solve math-like problems using logical rules to check which path to take next in a tree of LLM interactions. Our work performs semantic exploration for more general problem statements and allows the end-user to
not only decide on which path to explore next, but also learn from known alternative approaches to the same problem.

The importance of data cleaning for general LLM benchmark performance
was studied in~\cite{chen2024}. The cleaning strategies we propose are
not only syntactical but also semantic and include human alignment.
The effect of temperature on creativity was studied in~\cite{peeperkorn2024}, which concluded that there is only a weak connection and more than one parameter is needed to make LLMs assist with creativity tasks. This is consistent with the findings of our work, where we design a chatbot to guide an innovator through iterative ideation grounded in related work and constrained by domain-specific training. High diversity 
of generated ideas is desirable and use of off-the-shelf pre-trained LLMs could lead to design fixation, as shown in
~\cite{meincke2024,wadinambiarachchi2024}. We show that we are able to maintain high
diversity throughout our semantic navigation and data cleaning techniques. 

None of the literature cited above studied the use of locally-hosted LLMs of modest size, fine-tuned on an organization's proprietary internal data, for domain-constrained idea exploration and problem-to-solution and solution-to-problem space semantic traversal. Moreover, to the best of our knowledge, automated data cleaning using coherence, relevance and human alignment to improve generations to organizational knowledge has not been investigated before.

\section{Model}\label{sec:model}
\subsection{Problem and Solution Statement spaces}
We begin with the usual \emph{embedding} mapping~\cite{arsanjani2023} from text (usually at the word or character level) to a dense real vector of high dimension, e.g., $1536$ in the GPT-3.5 \texttt{text-embedding-3-small} embedding. Embeddings of problem (respectively solution) statements live in the problem (solution) statement space $\wp$ ($\mathcal{S}$) that is a subset of the cartesian product of as many embedding spaces as the maximum number of text segments that a problem (solution) statement text is split up into for embedding. Note that even with the same embedding used for problem and solution statements, $\dim \wp \neq \dim \mathcal{S}$ because the respective maximum numbers of text segments above are different for problem and solution statements.

\begin{figure}[htp]
        \centering
                \includegraphics[width=\textwidth]{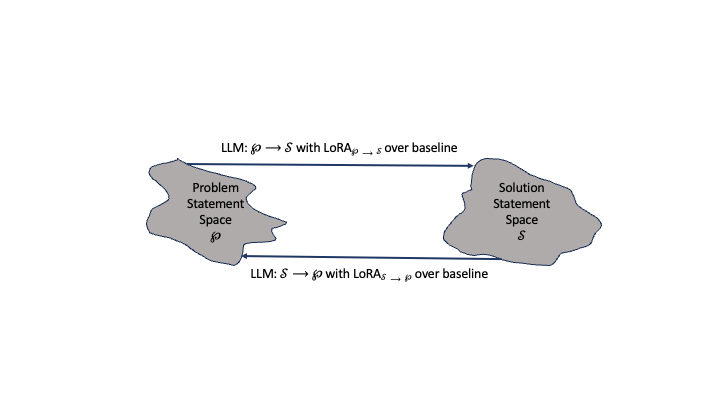}
	\caption{The pair of LLM-based mappings between the spaces of problem statements and solution statements.}
	\label{fig:llmprobsoln}
\end{figure}

We model our processing pipeline in terms of a pair of mappings, one from $\wp$ to $\mathcal{S}$, and the other from $\mathcal{S}$ to $\wp$, as shown in Fig.~\ref{fig:llmprobsoln}.  Note that these two mappings cannot be inverses of one another as $\dim \wp \neq \dim \mathcal{S}$.

\subsection{Mapping between Statement spaces using LLMs}
The mapping from, say, $\wp$ to $\mathcal{S}$ is done using a Large Language Model (LLM) implemented using the so-called \emph{Transformer} architecture~\cite{vaswani2017}. Although a baseline LLM will perform a mapping from one embedding space to another, its output is not tailored to the problem-statement-to-solution-statement mapping use case that is of interest to us here.  Therefore, we adapt the baseline pre-trained LLM to provide outputs better suited to this particular use case by \emph{fine tuning}~\cite{liu2022}, i.e., by further training it on a smaller set of (problem-statement, solution-statement) examples.  We further augment the fine tuning with Low-Rank Adaptation (LoRA)~\cite{hu2022}, a low-rank update matrix specific to the use case that is simply added to the Transformer matrix of the baseline LLM at the time of inference.

For our specific use case, we use a specific baseline LLM and maintain two LoRA updates, one for the mapping from $\wp$ to $\mathcal{S}$, the other for the mapping from $\mathcal{S}$ to $\wp$, as shown in Fig.~\ref{fig:llmprobsoln}.

\subsection{Exploring the Problem Statement spaces}
\label{sec:explore_statement_space}
\begin{figure}[htp]
        \centering
                \includegraphics[width=\textwidth]{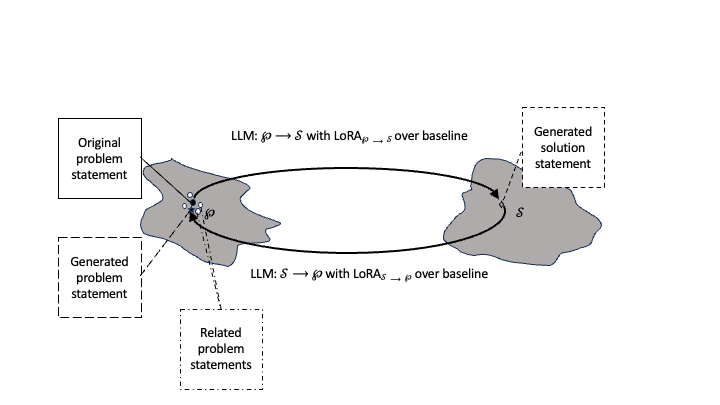}
	\caption{Exploring the Problem Statement space by finding related known problem statements and generating a new problem statement using the two LLM + LoRA mappings of Fig.~\ref{fig:llmprobsoln}.}
	\label{fig:llmprobsoln2}
\end{figure}

We want an ``Ideation Assistant'' tool that starts with an incomplete or unclear problem statement as input and outputs related problem statements with the intention of sparking creative ideation in the (human) user of the tool.  These related problem statement embeddings are points in $\wp$ that are ``near'' (by some metric) to the input problem statement, and are obtained in two ways, which we call \emph{selection} (of problem statements from a knowledge base of known problem-solution statement pairs that are semantically related to the input problem statement, e.g., the $4$ known problem statements whose embeddings have the highest cosine similarity to the input problem statement embedding\footnote{This operation is entirely performed in $\wp$ and neither LLM mapping in Fig.~\ref{fig:llmprobsoln} is applied.}) and \emph{sampling} (in $\wp$ and $\mathcal{S}$ by generating a new problem statement and mapping it to a solution statement).  This is shown in Fig.~\ref{fig:llmprobsoln2}, where semantic similarity is represented by geometric proximity, the original input problem statement is represented by a black dot and denoted $P_o$, and the $4$ related known problem statements are represented by the white dots in $\wp$.

\subsection{Wider exploration of the Problem Statement space}
The sequence of steps described in Sec.~\ref{sec:explore_statement_space} yields one newly-generated problem statement for each input problem statement, along with several known problem statements, all of which are related to the input problem statement.  This is useful not only for generating new problem statements, but also as a search tool for known problem statements that are related to the input problem statement.  Note also that we can pair the generated solution statement (obtained from the forward LLM + LoRA mapping applied to the original input problem statement) with the generated problem statement (obtained from the reverse LLM + LoRA mapping applied to the generated solution statement) and archive the generated (problem, solution) statement pair for further exploration if needed.\footnote{For example, the generated (problem, solution) statement pair could be inserted into the enterprise's knowledge base of known (problem, solution) statement pairs after suitable curation by human experts.}

\begin{figure}[htp]
        \centering
                \includegraphics[width=\textwidth]{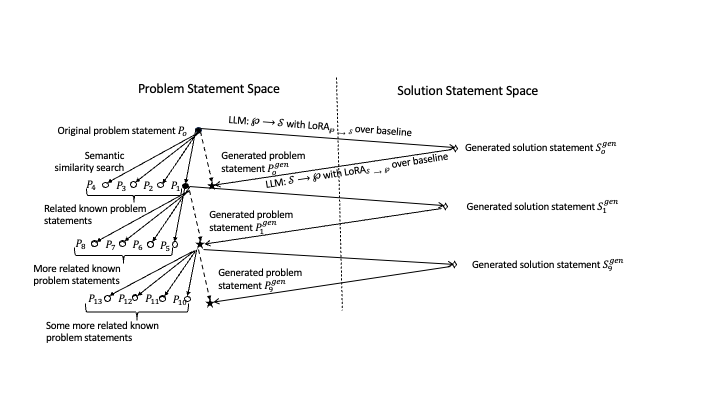}
	\caption{Illustrating a wide depth-first exploration of the Problem Statement space by iteratively applying the two LLM + LoRA mappings of Fig.~\ref{fig:llmprobsoln} to the newest generated problem statement, or alternatively, by searching for known problem statements related to one of the known problem statements discovered in the previous iteration.}
	\label{fig:llmprobsoln3}
\end{figure}

We may now reapply the procedure of Sec.~\ref{sec:explore_statement_space} taking the input problem statement as either the newly-generated problem statement or any of the known related problem statements discovered by the above step.  This corresponds to exploring the problem statement space $\wp$ through a depth-first tree-based traversal. 

The above procedure is illustrated in Fig.~\ref{fig:llmprobsoln3} (and formalized as an algorithm in Section~\ref{sec:motivation}).  From the original input problem statement $P_o$, we obtain the semantically related known problem statements $P_1, \dots, P_4$, as well as the new generated solution statement $S_o^{gen}$ and the new generated problem statement $P_o^{gen}$.  With reference to Fig.~\ref{fig:llmprobsoln2}, $P_o$ is the black dot, $S_o^{gen}$ the white diamond, and $P_o^{gen}$ the black star, and these icons are retained in Fig.~\ref{fig:llmprobsoln3} for consistency.  

As discussed above, at any stage, we could perform the same procedure on either one of the known related problem statements or on the new generated problem statement.  In Fig.~\ref{fig:llmprobsoln3}, we show what happens when we take known related problem statement $P_1$ as the input, yielding related known problem statements $P_5, \dots, P_8$, generated solution statement $S_1^{gen}$, and generated problem statement $P_1^{gen}$.  We also show what happens when, at the next stage, we take the new generated problem statement $P_1^{gen}$ as the input, thereby yielding related problem statements $P_{10}, \dots, P_{13}$, generated solution statement $S_9^{gen}$, and generated problem statement $P_9^{gen}$.  

\section{Motivation}\label{sec:motivation}
To motivate and validate our basic approach we use a problem-solution dataset extracted
using the OpenAI API and a public list of company names.

From a list of the top 400 software companies in terms of revenue 
we used the OpenAI API\footnote{with model gpt-3.5-turbo} and the following prompt:
\begin{verbatim}
Provide a short description of the problem the 
company _COMPANY_ solves and how it solves it 
separated by PROBLEM and SOLUTION headers without 
mentioning _COMPANY_ by name.
\end{verbatim}
where we replace the {\it COMPANY} tag with the name of the company for all companies in our list.
We found that we could get the LLM to provide valid solution and problem separations for 313 of the 400 companies tested, 
and that is thus our evaluation dataset.
Examples of these pairs can be found in Appendix~\ref{company}.

Given that we hope to extract a fairly complex NLP function of
mapping a problem to a solution (and a solution to a problem) with a small LLM  we opted for the approach of {\it fine tuning}
where we take a public foundation model trained on a large public corpus and then fine-tune it with a small set of examples of
prompt and expected output pairs. As foundation model we chose `Pythisa-Chat-Base-7B`\footnote{https://huggingface.co/togethercomputer/Pythia-Chat-Base-7B}. 
We train both the problem to solution and solution to problem mappings with the LoRA~\cite{hu2022} approach. Each mapping
is stored in a separate adapter that can be invoked and loaded on demand depending on which transformation to reproduce.
Each direction took about 19 minutes to train on a single GPU with this setup. Both adapters can also be pre-loaded into memory
to make inference efficient.

We also record the problems in a vector database (Redis) which allows for efficient retrieval of nearest-neighbor searches based
on publicly available embeddings\footnote{https://huggingface.co/sentence-transformers/all-mpnet-base-v2}.

Table~\ref{T:sim} shows the average and standard deviation of a test split withheld from the training data and
used in predictions (10 mappings) for pairwise
cosine similarity (ground truth versus generated), and Table~\ref{T:dist} shows the edit distance (Levenshtein) for
the same predictions (generations). 
The following predictors were used in this test:
\begin{itemize}
  \item{{\bf LoRA0.1}. Our fine tuned model with temperature set to $0.1$.} 
  \item{{\bf LoRA1.0}. Our fine tuned model with temperature set to $1.0$.} 
  \item{{\bf Random}. We pick a random problem or solution (depending on what is predicted) from the training data as the prediction.} 
  \item{{\bf OAI}. We feed the same prompt as was used for fine-tuning into the OpenAI API (which was also used to create the dataset).} 
  \item{{\bf OAIT}. We enhance the prompt with a tag such as ``Describe a problem/solution`` to clarify what we want OpenAI to do.} 
\end{itemize}

\begin{table}[htbp]
	\caption{Fine Tuning Similarity. 303/313 train, 10/313 test.}
\begin{center}
\begin{tabular}{|l|c|c|c|c|c|}
\hline
	\textbf{Transformation} &  \multicolumn{5} {c|} {\bfseries Similarity $\mu\pm\sigma$} \\
	 & \textbf{LoRA0.1} & \textbf{LoRA1.0} & \textbf{Random} & \textbf{OAI} & \textbf{OAIT} \\
\hline
	Problem$\rightarrow$Solution & $.85\pm.06$ & $.80\pm.07$ & $.39\pm.12$ & $.77\pm.11$ & $.81\pm.08$ \\
\hline
	Solution$\rightarrow$Problem & $.78\pm.10$ & $.75\pm.12$ & $.23\pm.10$ & $.70\pm.10$ & $.74\pm.14$ \\
\hline
\end{tabular}
\label{T:sim}
\end{center}
\end{table}

\begin{table}[htbp]
	\caption{Fine Tuning Distance. 303/313 train, 10/313 test.}
\begin{center}
\begin{tabular}{|l|c|c|c|c|c|}
\hline
	\textbf{Transformation} &  \multicolumn{5} {c|} {\bfseries Distance $\mu\pm\sigma$} \\
	 & \textbf{LoRA0.1} & \textbf{LoRA1.0} & \textbf{Random} & \textbf{OAI} & \textbf{OAIT} \\
\hline
	Problem$\rightarrow$Solution & $.53\pm.02$ & $.53\pm.03$ & $.55\pm.02$ & $.63\pm.07$ & $.65\pm.08$\\
\hline
	Solution$\rightarrow$Problem & $.49\pm.08$ & $.51\pm.06$ & $.54\pm.02$ & $.73\pm.08$ & $.57\pm.06$ \\
\hline
\end{tabular}
\label{T:dist}
\end{center}
\end{table}

We note that our fine-tuned models are able to replicate the original ground-truth semantically but not
lexically, which is desirable in our case of innovation idea exploration, and problem statement
refinement. Furthermore, the more natural problem to solution mapping yields a better result than the reverse
mapping, but both mappings are significantly higher than the random mapping, showing the feasibility of
semantic traversal in both directions. Looking at the OAI results we note that the semantic similarity
is not very high until we tag the prompt with helpful directions on what to output, showcasing that our
fine tuning works. Even with tagged prompts the OAI model, which here was used to produce the
ground truth, and hence should yield near optimal results, is slightly outperformed at the low temperature level $0.1$.
In summary, this test showed that our model can maintain semantic fidelity when traversing the solution and 
problem spaces, and that temperature plays a role in the semantic deviation. 

Appendix~\ref{exploration} shows how semantic and lexical distances are impacted by temperature in an exploration. 

\section{Data Cleaning and Filtering}\label{sec:cleaning}
The first problem we encountered when trying to use our approach with an internal idea database
was that poor quality input lead to poor generations.
Problems may not be well defined or the solution is not
self-contained or just a placeholder. 

Some poor quality inputs were easy to detect with regexp-type
filtering, e.g. ``see attached document'' or ``this is a merge of
several other ideas''.  Many of the ideas had other useful content
so they were not dropped, but these sentences were dropped to avoid
the LLM producing text referring to some non-existent external reference.
We also removed references to names of people, urls, special symbols and spurious newlines.

All these techniques we refer to as basic cleaning. In addition we performed
semantic filtering as well designed to exclude entire records from the training data
based on quality scores. The following three semantic filters were used:
\begin{itemize}
	\item{{\bf Relevance.} We compute the semantic (embedding) distance between 
		each sentence in the output and the prompt and take the complement of the average ($1-\mu$).}
	\item{{\bf Coherence.} We compute the semantic (embedding) distance between 
		each consecutive sentence in the output and take the complement of the average ($1-\mu$).}
	\item{{\bf Human alignment.} We pick a random output from the input and generate a prompt with our reverse mapper, then generate new alternative novel 
		outputs that are rated by human evaluators. About 600 such solutions and problems were generated in total. They were then evaluated by 9 human expert innovators\footnote{none of these innovators participated in the subsequent user study}
		and used to create an RLHF reward model. On average each expert evaluated about 20 solution generations and 20 problem generations, with a total of 360 votes cast. The real inputs were then evaluated using this reward model to asses human alignment.}
\end{itemize}
Using these filtering techniques we created three data sets that were separately trained to create different LoRA adapters to use for generation: 
\begin{itemize}
	\item{{\bf Default.} Takes the 600 ideas and applies basic cleaning.} 
	\item{{\bf RelCohTop500.} Takes the Default data set then filters out the bottom 100 based on the linear combination of relevance and coherence scores.} 
	\item{{\bf RewardTop400.} Takes the RelCohTop500 data set and then filters out the bottom 100 based on the human alignment score.} 
\end{itemize}

\begin{figure}[htp]
        \centering
         \includegraphics[width=2in]{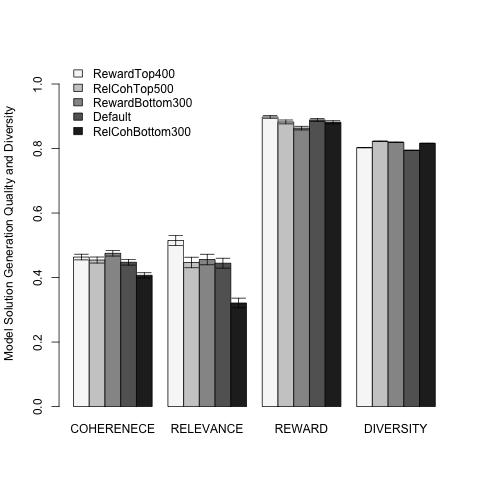}
         \includegraphics[width=2in]{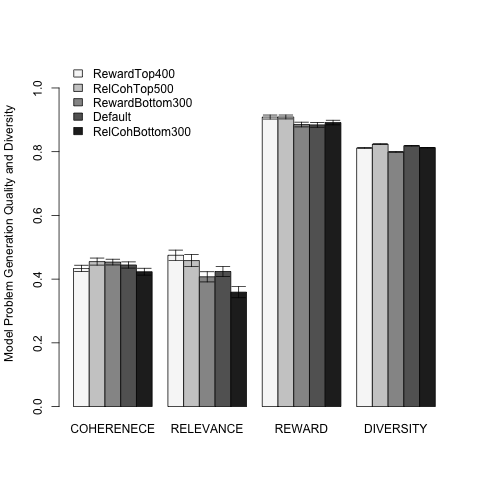}
	\caption{Evaluation of input filtering conditions.}
	\label{usermodeleval}
\end{figure}

We then generated new output with these three adapters and evaluated the three semantic scores both for solution and problem generation in Figure~\ref{usermodeleval},
together with some additional benchmarks RelCohBottom300 (worst performing records using the relevance and coherence score) and RewardBottom300 (worst
performing records using human alignment scores). We see that the filtering carries over to the generation as expected, i.e., we can improve the relevance, coherence and
human alignment of the generated output but simply removing records with low scores in those metrics from the input without negatively impacting diversity.
Diversity here is just the pairwise embedding vector distance between all inputs and outputs. Each model was evaluated with 300 generations.

These three LoRA adapters were served to different users as conditions in a between-subject user study described next.

\section{User Study}\label{sec:userstudy}
We conducted a study with human subjects to assess
the usefulness of our ideation assistant for ideation tasks.

\subsection{Study Design}
\label{sec:study_design}
The user study had two goals: to assess whether users noticed a difference in brainstorming while using the tool under different conditions and to evaluate their response to a Slack bot based LLM interface. The first tested the LLM tuned with the existing idea database under three conditions (Default, RelCohTop500 or RewardTop400), while the second gathered user experience feedback critical for future work. The study had two phases, each followed by a custom survey with overlapping questions to track changes in user perceptions. Voting features were added to let users rate the quality of tool responses.
\begin{itemize}
\item{{\bf Phase 1.} 15 innovators were recruited and given an initial personal 15-min tutorial. Each user was given the same problem statement prompt and asked to evaluate and explore the generated solution. All the solutions in this first phase were generated using the same Default model for all users. The subject could vote thumbs down, thumbs up or tada (very good) to give feedback on the generated responses. This tutorial approach ensured all users received consistent training in the platform and that differing rates of climbing the learning curve would not be a factor the platform's perceived performance.}

\item{{\bf Phase 2.} In Phase 2, 15 users were split into three groups of five, each assigned to a Slack channel with one of three LLM fine-tuning conditions. Users explored their own problem statements, with each group using a different model (Default, RelCohTop500, or RewardTop400). They had one week to explore at their own pace and were regularly prompted to engage, ensuring enough interaction to support the Phase 2 survey and adequately test the conditions.} 
\end{itemize}

The instrumentation allowed for the user experience (UX) to be assessed across multiple dimensions, assess which of the test conditions drove the greatest improvements in perceived performance and to gather valuable insights on what shape the next iteration of the platform should take.  The analysis leverages the extensive user data, both qualitative and quantitative, gathered during the user study phases.

\subsection{AI Assistant Tool}
\label{sec:ai_assistant_tool}
We developed a slack bot that can be triggered by adding a tag to message posts,
such as \%idea (for a generation against the internal idea database) or \%pitch (for a generation against the public dataset).
We generate an immediate response to confirm that the idea is being explored by
the assistant in a thread to the post. After about $20\,\text{s}$ we return the generated
solution in the same thread as well as related ideas, potential tags used in those
ideas, as well as a series of buttons for further exploration.

\begin{figure}[htp]
        \centering
         \includegraphics[width=5in]{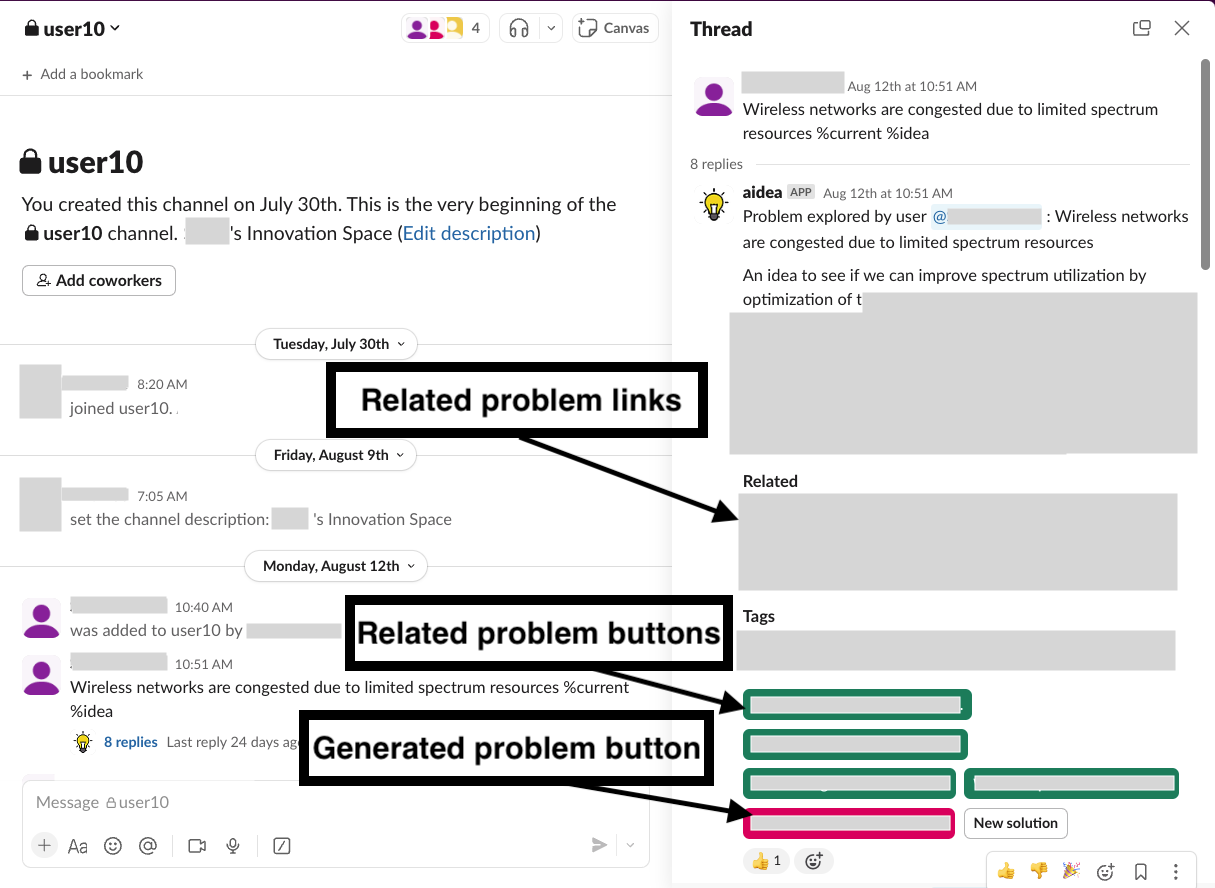}
	\caption{Slack Ideation Assistant UI.}
	\label{ui}
\end{figure}

The buttons have the following functions: {\bf green}, allows problems from related ideas
to be used as prompts to generate new solutions, {\bf red} a problem that was generated
from the generated solution that could be explored to generate a new solution, sometimes
referred to as a refined problem statement, {\bf white} a new solution can be generated
from the same problem statement that was originally entered. A sample of the UI
and some interactions against the public data set can be seen in Figure~\ref{ui}.

\subsection{Vote Analysis}
\label{sec:vote_analysis}
\begin{figure}[htp]
        \centering
         \includegraphics[width=4in]{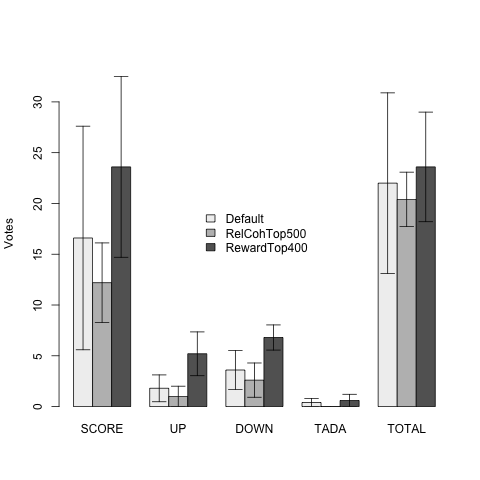}
	\caption{Votes for different conditions.}
	\label{uservoteeval}
\end{figure}

We first looked at explicit feedback in terms of thumbs up, thumbs down and ``tada'' emoticons tagged to the generations by the subjects.
We compute an overall score per condition (or model) as follows:
\begin{equation}
	s(p) = g(p) - e_{td}(p) + 2 e_{tu}(p) + 3 e_{ta}(p) 
\end{equation}
\begin{equation}
	s = s(2)-s(1)
\end{equation}
where $s(p)$ is the score from the phase $p$ votes, $g$ is the total number of generations $e_{td}$ are number of thumbs down votes,
$e_{tu}$ number of thumbs up votes, $e_{ta}$ number of tada votes. The score $s$ can be interpreted as the increase in vote score between the two phases.
The second phase score is expected to by higher as a longer time was used
for explorations, and it is not possible to have negative scores. Figure~\ref{uservoteeval} shows that the \textit{RewardTop400} model
performed best but surprisingly the \textit{RelCohTop500} model did not score higher than \textit{Default}. The error bars are standard errors.
Given the presence of some outliers, we computed a one-sided t-test removing the top and bottom scores and found that \textit{RewardTop400}
outscored \textit{Default} at a $10$\% confidence level (p-value $0.1$).

\subsection{Net Promoter Score Analysis}\label{sec:surveyanalysis}

Net Promoter Score (NPS) is a metric to measure user satisfaction by asking how likely users would use or recommend a company/product. In our case here, we analyze the NPS score $\Delta$ between phase 1 and phase 2 of the user study to look into how users change in their scoring. The question asked in both the initial and final user study survey was: ``How likely are you to use AI-deation for a future brainstorm with yourself?'' Scores are categorized into Promoters (voted 5 and mapped to +1), Passively Satisfied (voted 4 and mapped to 0), and Detractors (voted 0-3 and mapped to -1). The raw NPS score is then calculated by subtracting the percentage of detractors from promoters, providing clear and actionable insight into user sentiment.

\begin{figure}[H]   
        \centering
         \includegraphics[width=6in]{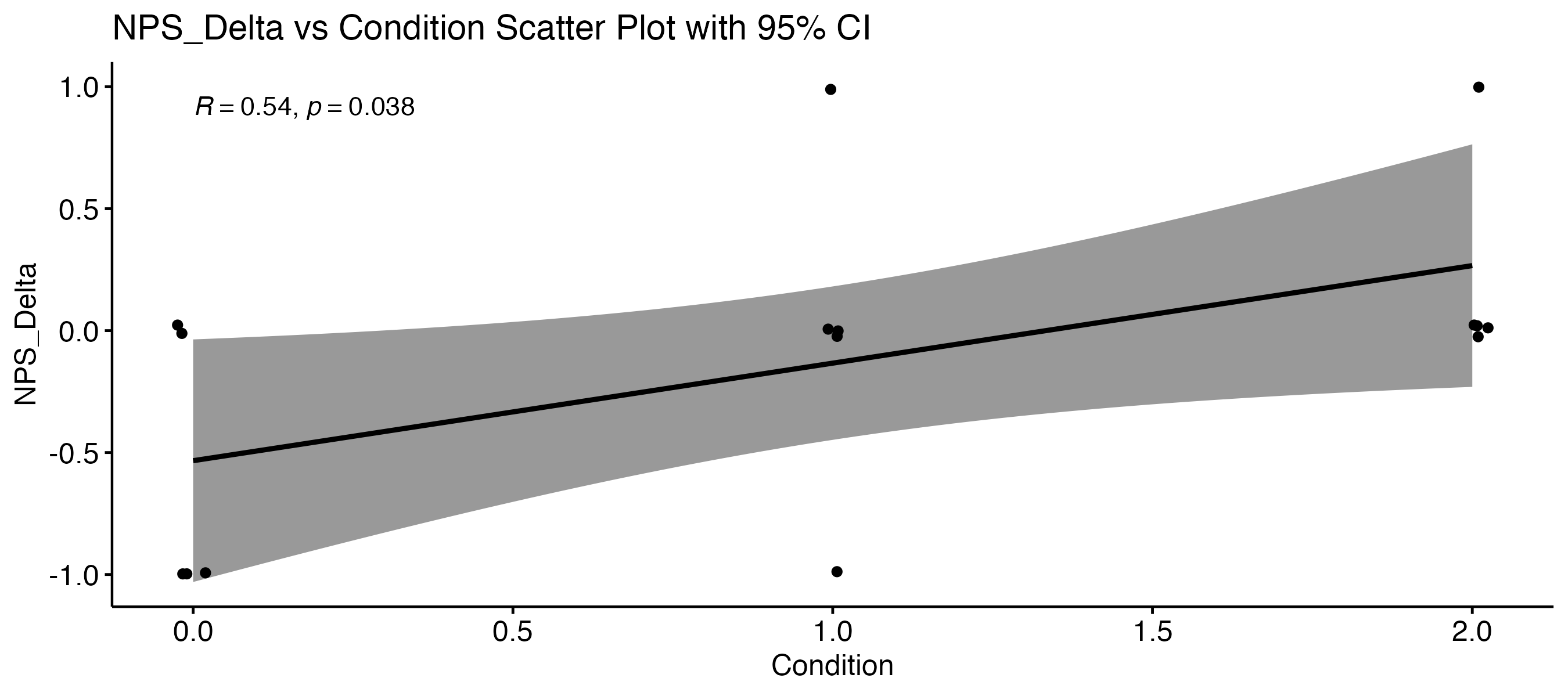}
	\caption{NPS $\Delta$ vs Condition scatterplot with $95\%$ Confidence Interval.}
        \label{NPSdeltaconditionscatter}
  \centering
 \end{figure}
Figure~\ref{NPSdeltaconditionscatter} reveals a strong positive correlation between NPS $\Delta$ and Conditions. The correlation coefficient of $0.54$ and p-value of $0.038$ from the Spearman Correlation test suggests that each succeeding condition tends to have a higher NPS $\Delta$ than that of the prior condition. 

\begin{figure}[H]
        \centering
         \includegraphics[width=6in]{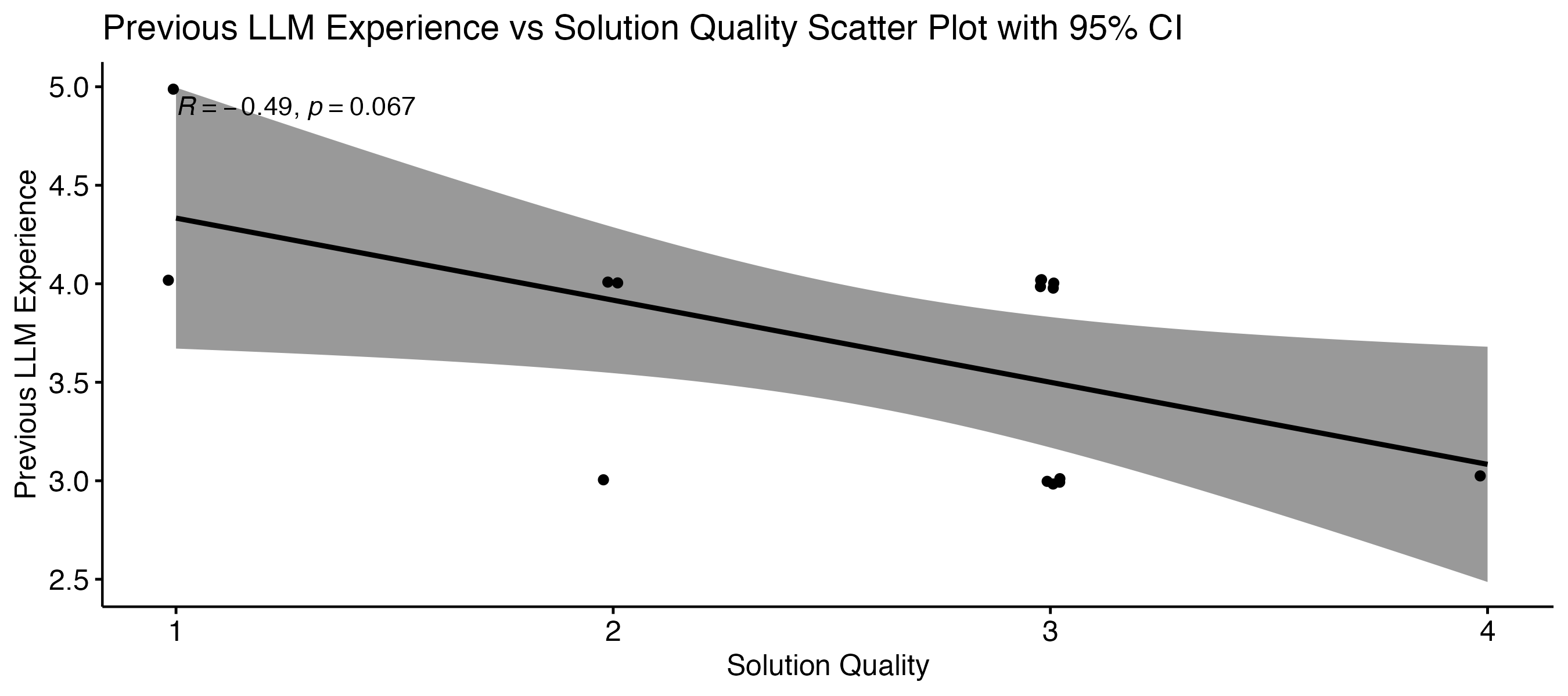}
	\caption{Previous LLM Experience vs Solution Quality scatterplot.}
	\label{LLMexperiencequality}
\end{figure}
Figure~\ref{LLMexperiencequality} shows a strong negative correlation between previous LLM experience and the users' perceived solution quality. The correlation coefficient of $-0.49$ and p-value of $0.067$ from the Spearman Correlation test suggests that the more past experience a user has had with LLMs and generative AI, the lower they tend to rate the solution quality. \newline
This then brings up the question of whether or not previous LLM experience is the factor determining user satisfaction as measured by NPS $\Delta$ or if it is due to the differing conditions. By holding the past LLM experience variable constant and analyzing the relationship between conditions and NPS $\Delta$ (NPS $\Delta$ ~ Condition + Past LLM Experience), we obtain a p-value of $0.052$. This further investigation supports the correlation between conditions and NPS $\Delta$, with past LLM experience not being a significant confounding variable. 

\subsection{Semantic Traversal Analysis}
\label{sec:semantic_traversal_analysis}
To analyze the value of our semantic traversal model exposed in the UI through related problem statements,
generated problem statements and the `new solution' re-generation buttons we compared top level prompts
to explore solutions versus the number of explorations through buttons inside a slack thread for the original
problem statement. We found that across all our 15 user we had 58 root explorations
and 119 threaded explorations (not including the root explorations). That is, $2.1$ times as many
generations were performed through our semantic navigation features than through prompts.
A one-sided t-test of the hypothesis that threaded exploration was higher showed a p-value of
$0.01$ across the study participants. 

One survey question asked the participants to rank the 4 exploration modes from most to least preferred (see Appendix~\ref{surveyquestions} phase 2, question 4-8).
The rankings were consistent across the different
conditions and the consensus ranking in the order of most to least preferred was:
\begin{itemize}
\item{{\bf 1} - generating a new solution for the same prompt (sum of ranks across conditions: 26)}
\item{{\bf 2} - generating a new solution from a problem generated from the generated solution (37)}  
\item{{\bf 3} - generating a new solution from a related project problem (43)}  
\item{{\bf 4} - entering new prompts (44)}  
\end{itemize}
Lowest rank of 1-3 versus the rank of 4 for a Welch two-sample t-test for alternative hypothesis 
``true difference in means is less than 0'', resulted in a p-value of $6 \times 10^{-6}$. 
This result further supports our hypothesis that semantic navigation options were preferred to the traditional prompt
editing explorations.

\section{Implementation Notes}\label{implementation}
We implemented the solution using the Slack Websocket and Web Client API hosted on GPU servers
with $3$ NVIDIA $48$GB RAM GPUs across two VMs serving $7$B parameter LLM models. 
Requests from Slack were load-balanced
across the GPUs and directed to servers hosting the model that the user
was assigned to. For added concurrency, each model was loaded in two servers
across two GPUs. Problem-to-Solution generation can be done concurrently
with Solution-to-Problem generation on the same server. A generation
without contention takes about $10-15$ seconds. All code was implemented 
with pytorch and CUDA, and the Slack front-end communicated with the model
inference servers using a REST API both within and across our GPU VMs.
To improve speed and contain randomness we limit the problem generations
to $50$ tokens and the solution generations to $150$ tokens. Typically there
are about $4$ tokens used for each $3$ words.

\section{Conclusion}\label{sec:conclusion}
Our user study confirmed the importance of input data quality when fine tuning
LLMs for creative output. Careful selection of the highest quality input to
use for training using quality metrics such as relevance, coherence and
human alignment not only transferred these properties to the generations
as well, but lead to an improved user experience.

Semantic navigation as a model for solution and problem exploration during
ideation allowed for controlled, domain-constrained discovery of related
work as well as novel solution and problem creation by means of LoRA
fine tuning and text embedding distance nearest neighbor search.

The tool proved useful in enabling individual brainstormers to harness the organizational memory during their ideation workflows.  Interestingly, there was no consistent toolchain in the participants' ideation workflows and, even with expressed concerns over the UI, most participants found value in integrating a tool.

Temperature in an LLM generation could be used to effectively control the size of the semantic domain explored, but more work is needed to determine how the extra randomness impacts the user experience. From a qualitative analysis of user study participant feedback we did see indicators pointing toward the random solution quality both bothering and intriguing participants.

The subjects of the user study were volunteers solicited from employees of the authors' company.  Users were not compensated for their participation.  All survey results, including verbatims, were anonymized.  Participation was in alignment with both company and ACM policy on human subjects.

\bibliographystyle{plain}
\bibliography{related}

\begin{thebibliography}{10}

\bibitem{arsanjani2023}
Ali Arsanjani.
\newblock {The Evolution of Text Embeddings}, 2023.

\bibitem{chakrabarty2023}
Tuhin Chakrabarty, Vishakh Padmakumar, Faeze Brahman, and Smaranda Muresan.
\newblock {Creativity Support in the Age of Large Language Models: An Empirical
  Study Involving Emerging Writers}.
\newblock {\em arXiv preprint arXiv:2309.12570}, 2023.

\bibitem{chen2024}
Daoyuan Chen, Yilun Huang, Zhijian Ma, Hesen Chen, Xuchen Pan, Ce~Ge, Dawei
  Gao, Yuexiang Xie, Zhaoyang Liu, Jinyang Gao, et~al.
\newblock Data-juicer: A one-stop data processing system for large language
  models.
\newblock In {\em Companion of the 2024 International Conference on Management
  of Data}, pages 120--134, 2024.

\bibitem{di2022}
Giulia Di~Fede, Davide Rocchesso, Steven~P Dow, and Salvatore Andolina.
\newblock {The Idea Machine: LLM-based Expansion, Rewriting, Combination, and
  Suggestion of Ideas}.
\newblock In {\em Proceedings of the 14th Conference on Creativity and
  Cognition}, pages 623--627, 2022.

\bibitem{diehl1987productivity}
Michael Diehl and Wolfgang Stroebe.
\newblock Productivity loss in brainstorming groups: Toward the solution of a
  riddle.
\newblock {\em Journal of personality and social psychology}, 53(3):497, 1987.

\bibitem{hu2022}
Edward~J Hu, Yelong Shen, Phillip Wallis, Zeyuan Allen-Zhu, Yuanzhi Li, Shean
  Wang, Lu~Wang, and Weizhu Chen.
\newblock {Lo{RA}: Low-Rank Adaptation of Large Language Models}.
\newblock In {\em International Conference on Learning Representations}, 2022.

\bibitem{hwang2021}
Angel Hsing-Chi Hwang and Andrea~Stevenson Won.
\newblock Ideabot: investigating social facilitation in human-machine team
  creativity.
\newblock In {\em Proceedings of the 2021 CHI Conference on Human Factors in
  Computing Systems}, pages 1--16, 2021.

\bibitem{koch2019}
Janin Koch, Andr\'{e}s Lucero, Lena Hegemann, and Antti Oulasvirta.
\newblock {May AI? Design Ideation with Cooperative Contextual Bandits}.
\newblock In {\em Proceedings of the 2019 CHI Conference on Human Factors in
  Computing Systems}, CHI '19, page 1–12, New York, NY, USA, 2019.
  Association for Computing Machinery.

\bibitem{liu2022}
Haokun Liu, Derek Tam, Muqeeth Mohammed, Jay Mohta, Tenghao Huang, Mohit
  Bansal, and Colin Raffel.
\newblock Few-shot parameter-efficient fine-tuning is better and cheaper than
  in-context learning.
\newblock In Alice~H. Oh, Alekh Agarwal, Danielle Belgrave, and Kyunghyun Cho,
  editors, {\em Advances in Neural Information Processing Systems}, 2022.

\bibitem{liu2023}
Yiren Liu, Si~Chen, Haocong Cheng, Mengxia Yu, Xiao Ran, Andrew Mo, Yiliu Tang,
  and Yun Huang.
\newblock {How AI Processing Delays Foster Creativity: Exploring Research
  Question Co-Creation with an LLM-based Agent}.
\newblock {\em arXiv preprint arXiv:2310.06155}, 2023.

\bibitem{long2023}
Jieyi Long.
\newblock Large language model guided tree-of-thought.
\newblock {\em arXiv preprint arXiv:2305.08291}, 2023.

\bibitem{meincke2024}
Lennart Meincke, Ethan~R Mollick, and Christian Terwiesch.
\newblock Prompting diverse ideas: Increasing ai idea variance.
\newblock {\em arXiv preprint arXiv:2402.01727}, 2024.

\bibitem{ouyang2022}
Long Ouyang, Jeffrey Wu, Xu~Jiang, Diogo Almeida, Carroll Wainwright, Pamela
  Mishkin, Chong Zhang, Sandhini Agarwal, Katarina Slama, Alex Ray, et~al.
\newblock Training language models to follow instructions with human feedback.
\newblock {\em Advances in neural information processing systems},
  35:27730--27744, 2022.

\bibitem{peeperkorn2024}
Max Peeperkorn, Tom Kouwenhoven, Dan Brown, and Anna Jordanous.
\newblock Is temperature the creativity parameter of large language models?
\newblock {\em arXiv preprint arXiv:2405.00492}, 2024.

\bibitem{shneiderman2002}
Ben Shneiderman.
\newblock Creativity support tools.
\newblock {\em Communications of the ACM}, 45(10):116--120, 2002.

\bibitem{si2024}
Chenglei Si, Diyi Yang, and Tatsunori Hashimoto.
\newblock {Can LLMs Generate Novel Research Ideas? A Large-Scale Human Study
  with 100+ NLP Researchers}.
\newblock {\em arXiv preprint arXiv:2409.04109}, 2024.

\bibitem{sutton1996}
Robert Sutton and Andrew Hargadon.
\newblock Brainstorming groups in context: Effectiveness in a product design
  firm.
\newblock {\em Administrative Science Quarterly}, 41:685--718, 12 1996.

\bibitem{vaswani2017}
Ashish Vaswani, Noam Shazeer, Niki Parmar, Jakob Uszkoreit, Llion Jones,
  Aidan~N Gomez, \L~ukasz Kaiser, and Illia Polosukhin.
\newblock {Attention is All you Need}.
\newblock In I.~Guyon, U.~Von Luxburg, S.~Bengio, H.~Wallach, R.~Fergus,
  S.~Vishwanathan, and R.~Garnett, editors, {\em Advances in Neural Information
  Processing Systems}, volume~30. Curran Associates, Inc., 2017.

\bibitem{wadinambiarachchi2024}
Samangi Wadinambiarachchi, Ryan~M Kelly, Saumya Pareek, Qiushi Zhou, and
  Eduardo Velloso.
\newblock The effects of generative ai on design fixation and divergent
  thinking.
\newblock In {\em Proceedings of the CHI Conference on Human Factors in
  Computing Systems}, pages 1--18, 2024.

\bibitem{yu2011}
Lixiu Yu and Jeffrey~V Nickerson.
\newblock {Cooks or cobblers? Crowd creativity through combination}.
\newblock In {\em Proceedings of the SIGCHI conference on human factors in
  computing systems}, pages 1393--1402, 2011.

\end{thebibliography}

\newpage
\appendix
\section{Company Dataset LLM Prompt and Output Examples}\label{company}
\begin{table}[htbp]
        \caption{Prompts and responses from OpenAI API.}
\tiny
\begin{center}
\begin{tabular}{|p{.99\linewidth}|}
\hline
Provide a short description of the problem the company {\bf Apple} solves and how it solves it separated by PROBLEM and SOLUTION headers without mentioning {\bf Apple} by name.\\
{\bf PROBLEM:} \\
The problem this company tackles is the need for accessible and user-friendly technology. Many individuals struggle with complicated devices and find it challenging to navigate through various applications and features. Additionally, there is a demand for seamless integration between different devices and systems.\\
{\bf SOLUTION:} \\
The company offers a range of products and services that provide a simplified and intuitive user experience. Their devices are designed with a user-friendly interface, making it easier for individuals to interact with technology. Furthermore, their ecosystem allows for seamless integration among different devices, providing a cohesive and streamlined experience across multiple platforms. Through their innovative solutions, the company aims to bridge the gap between technology and its users, ensuring that anyone can access and utilize their products effortlessly. \\ 
\hline
Provide a short description of the problem the company {\bf Alphabet (Google)} solves and how it solves it separated by PROBLEM and SOLUTION headers without mentioning {\bf Alphabet (Google)} by name. \\
{\bf PROBLEM:} \\
Individuals and businesses often struggle with finding and organizing information from a vast amount of sources on the internet. The complex and ever-expanding nature of the online world poses a challenge in navigating and accessing relevant data efficiently. This can lead to information overload, wasted time, and difficulty in decision-making.\\
{\bf SOLUTION:} \\
The company has developed advanced algorithms and technologies to index, analyze, and present information from the internet in a fast and accurate manner. Through its search engine, the company enables users to effortlessly search for and retrieve relevant information from billions of web pages. By employing various ranking factors and machine learning techniques, the solution provides users with the most helpful and contextually accurate search results. Additionally, the company offers various additional services and tools to further assist individuals and businesses in their information management, productivity, and communication needs.\\ 
\hline
Provide a short description of the problem the company {\bf Microsoft} solves and how it solves it separated by PROBLEM and SOLUTION headers without mentioning {\bf Microsoft} by name. \\
{\bf PROBLEM:} \\
Many people and businesses struggle with using and managing their computer systems efficiently. They often face challenges such as disorganization, security vulnerabilities, lack of communication, and compatibility issues.\\
{\bf SOLUTION:} \\
To address these problems, a technology company offers a wide range of software products and services that aim to streamline computer operations, enhance productivity, foster collaboration, and secure information. Their solutions include operating systems, productivity tools, communications platforms, and cloud services. By developing and delivering innovative software solutions, the company helps users and organizations address their computer-related challenges and improve their overall efficiency and effectiveness.\\
\hline
Provide a short description of the problem the company {\bf IBM} solves and how it solves it separated by PROBLEM and SOLUTION headers without mentioning {\bf IBM} by name. \\
{\bf PROBLEM:} \\
Many organizations struggle with effectively managing and analyzing vast amounts of data generated in today's digital world. Traditional methods of data handling and processing often prove inefficient, leading to missed opportunities for growth, insights, and cost savings. Additionally, the complexity and variety of data sources further compound these challenges and hinder decision-making processes.\\
{\bf SOLUTION:} \\
To address this problem, a company offers a comprehensive data management and analytics solution. Their solution leverages advanced technologies, such as artificial intelligence and machine learning, to streamline the organization's data operations. The platform integrates a wide range of data sources, providing a centralized and unified view of the organization's data assets. By employing advanced analytics algorithms, the solution helps businesses uncover valuable insights and patterns within their data. These insights empower organizations to make data-driven decisions and establish a competitive edge in their industry. Furthermore, the solution enables efficient data governance, privacy, and security measures to ensure compliance with regulations and protect sensitive information. Overall, the company's solution enables organizations to harness the power of their data, optimize operations, improve customer experiences, and drive innovation across various sectors. \\
\hline
Provide a short description of the problem the company {\bf Oracle} solves and how it solves it separated by PROBLEM and SOLUTION headers without mentioning {\bf Oracle} by name. \\
{\bf PROBLEM:} \\
Businesses generate a vast amount of data ranging from customer information to financial records. However, managing and organizing this data can be a challenging task. Additionally, as businesses grow and expand, the task of maintaining and integrating multiple databases from different departments becomes increasingly complex. This can lead to disjointed information, inefficiencies, and difficulties in making informed decisions in a timely manner.\\
{\bf SOLUTION:} \\
To overcome these challenges, a comprehensive database management system is required. Such a system enables businesses to efficiently store, organize, and retrieve data, while also ensuring its security and integrity. It provides a centralized platform that integrates various databases, allowing for seamless sharing of information across departments. Moreover, advanced analytics and reporting capabilities empower businesses to derive meaningful insights from their data, aiding in better decision-making processes. By offering a holistic solution for database management and analysis, this technology helps businesses streamline their operations, improve efficiency, and drive growth. \\
\hline
Provide a short description of the problem the company {\bf Salesforce} solves and how it solves it separated by PROBLEM and SOLUTION headers without mentioning {\bf Salesforce} by name. \\
{\bf PROBLEM:} \\
Companies face difficulties in managing their customer data, leads, and sales processes efficiently. Sales teams often struggle with disorganized information, lack of collaboration, and difficulty in tracking progress and pipelines. This can lead to missed opportunities, poor customer experiences, and a decrease in overall sales effectiveness.\\
{\bf SOLUTION:} \\
A company provides a comprehensive software solution that solves these challenges by offering a centralized platform for managing customer data, leads, and sales processes. This solution allows companies to track and analyze their sales pipelines, collaborate with team members, automate repetitive tasks, and streamline the overall sales process. With easy access to real-time data and personalized insights, sales teams can make informed decisions, enhance customer interactions, and increase their overall productivity. This software solution offers scalability and customization options, catering to the unique needs of every business, and ultimately improving their sales efficiency and driving growth.\\
\hline
Provide a short description of the problem the company {\bf Automatic Data Processing} solves and how it solves it separated by PROBLEM and SOLUTION headers without mentioning {\bf Automatic Data Processing} by name. \\
{\bf PROBLEM:} \\
Companies face challenges in managing and processing their employee data, payroll, and other human resources tasks efficiently. Traditional manual methods can result in errors, delays, and compliance issues.\\
{\bf SOLUTION:} \\
The company provides a comprehensive and automated platform that streamlines and simplifies employee management and human resources processes. Their solution integrates payroll, benefits administration, time and attendance tracking, and other HR functions into a single system. This eliminates manual data entry, reduces errors, ensures compliance with regulations, and enhances overall efficiency.\\
\hline
\end{tabular}
\end{center}
\end{table}

\newpage
\section{Problem to Solution Mapping Examples}\label{solution}
\begin{table}[htbp]
	\caption{Solutions generated from problem statements at different temperature levels.}
\tiny
\begin{center}
\begin{tabular}{|l|p{.45\linewidth}|p{.45\linewidth}|}
\hline
{\bf Temp} & {\bf Problem} & {\bf Generated Solution} \\
\hline
0.5 & The problem is that companies need to effectively manage and optimize their product development processes, including complex engineering tasks, collaboration across teams, and efficient deployment of resources. Traditional methods often result in inefficiencies, data fragmentation, and lack of visibility, hampering productivity and innovation. &   By leveraging advanced technologies, such as cloud-based platforms and artificial intelligence (AI), the company offers a comprehensive solution to address the problem of product development management. Their solution offers a suite of tools and services that streamline collaboration, automate workflows, and enable real-time monitoring and analysis of project data. By integrating various components like design tools, simulation, and engineering simulation, the solution enables companies to efficiently manage the entire product development lifecycle. Additionally, the platform provides a centralized hub for project stakeholders to track and manage tasks, assign responsibilities, and track progress. This enables a more coordinated, efficient, and data-driven approach to product development. \\
\hline
0.6 & Many companies struggle with effectively managing their software development processes. They often face challenges in coordinating tasks, tracking progress, and ensuring efficient collaboration among team members. This can lead to delays, miscommunication, and reduced productivity. &   By providing a comprehensive software development platform, this company enables businesses to effectively manage their software development processes. Their solution includes robust collaboration tools, task management systems, and integrated development environments. These tools streamline communication, promote collaboration, ensure timely completion of tasks, and drive overall efficiency in software development. By providing a central hub for team members to access and update information, the solution helps companies tackle the challenges of coordinating multiple tasks, tracking progress, and ensuring smooth collaboration. With its integrated development environments, businesses can easily prototype, code, and test their software, ultimately leading to better quality and faster time to market. \\
\hline
0.7 & Many companies struggle with effectively managing their software development projects. They often face challenges in terms of project scope, requirements gathering, development timeline, and coordination amongst team members. Due to these complexities, projects frequently experience delays, budget overruns, and poor product quality. &   The company offers a comprehensive software development platform that helps organizations streamline and optimize their software development processes. By providing robust tools, the platform assists teams in effectively managing project scope, requirements gathering, development timeline, and overall team coordination. This helps companies meet deadlines, minimize budget overruns, and deliver high-quality products with reduced risk and overall efficiency. With its comprehensive suite of tools and expertise, the company empowers organizations to developer innovative solutions efficiently and effectively. \\
\hline
0.8 & Many companies struggle with managing projects, tasks, and team collaboration efficiently. Traditional methods of tracking progress, communicating updates, and coordinating resources can be cumbersome and lead to confusion and delays. This lack of streamlined workflow can result in decreased productivity and hinder the overall success of a project. &   The company provides a comprehensive project management platform that integrates various tools and features to replace the traditional method of managing projects. Their platform enables team collaboration through real-time communication, task management, and project boards. It ensures accurate tracking of tasks, organizes critical documentation, and provides updates to stakeholders in a centralized location. With this solution, companies can streamline project management, reduce confusion, enhance collaboration, and achieve better efficiency. By reducing time wasted on administrative tasks and improving communication, this platform enhances productivity, reduces errors, and ultimately drives project success. 
The company's solution empowers teams to effectively manage their projects, stay informed, and make informed decisions. \\
\hline
0.9 & Many organizations struggle to effectively manage their projects and track progress. They find it challenging to collaborate and communicate efficiently, which leads to delayed deadlines, miscommunication, and ultimately hampers productivity. Additionally, accessing relevant project information and ensuring alignment across teams can be a cumbersome task. &   The company provides an integrated project management system that aids organizations in streamlining tasks, ensuring progress visibility, and fostering collaboration. Their software enables teams to efficiently mobilize resources, track progress, and monitor deliverables in real-time. It also allows for effective communication, allowing for real-time updates, comments, and reviews, ensuring clarity and transparency throughout the project lifecycle. By offering a user-friendly and comprehensive platform, the company enhances productivity and enables organizations to achieve their goals faster. \\
\hline
1.0 & Many organizations struggle with effective team collaboration and project management due to a lack of proper communication channels and disjointed workflows. Teams may find it difficult to share information, track progress, and coordinate efforts, leading to delays, misalignment, and inefficiencies. &   By leveraging innovative technology, a service-oriented company offers a comprehensive solution for effective team collaboration and project management. Their platform provides seamless communication channels, facilitating timely and secure chatting, voicing, and sending of various project-related documents. Additionally, their platform provides powerful tools such as project backlog interfaces, Kanipboxes, and join.me for visually organizing project teams and facilitating collaboration. Furthermore, their services aim to streamline workflows by automating and organizing project tasks, leading to higher efficiency and productivity. With this solution, organizations can ensure that their teams work cohesively, efficiently, and on the right track, thereby fostering greater success in achieving their goals. \\
\hline
1.1 & Many companies struggle with managing projects, tasks, and team collaboration efficiently. Traditional methods of tracking progress, communicating updates, and coordinating resources can be cumbersome and lead to confusion and delays. This lack of streamlined workflow can result in decreased productivity and hinder the overall success of a project. &   To address this problem, a company has developed a breakthrough software solution that enables efficient project management. This solution enables individuals and teams to organize projects, assign tasks, track progress, and share documents effortlessly. With its intuitive interface and robust features, this software automates and simplifies project management, significantly reducing friction and enhancing collaboration. It presents a clear overview of resources, upcoming deadlines, and progress updates, facilitating informed decision-making and smooth execution. With this solution, companies can confidently deliver better project outcomes while enhancing productivity and efficiencies. \\
\hline

\end{tabular}
\end{center}
\end{table}

\newpage
\section{Solution to Problem Mapping Examples}\label{problem}
\begin{table}[htbp]
	\caption{Problem statements generated from solutions at different temperature levels.}

\tiny
\begin{center}
\begin{tabular}{|l|p{.45\linewidth}|p{.45\linewidth}|}
\hline
{\bf Temp} & {\bf Solution} & {\bf Generated Problem} \\
\hline
0.5 &   The company provides a comprehensive project management solution that integrates various tools and features to tackle the challenges associated with managing multiple projects simultaneously. Their platform enables teams to create and track tasks, track progress, assign roles and permissions, create milestone check-ins, and communicate updates efficiently. This solution simplifies project management, enhances collaboration, and fosters efficient workflow, leading to increased productivity and success. By streamlining project management, the company empowers teams to effectively manage their projects and achieve their desired outcomes. & Many companies struggle with managing multiple projects and tasks simultaneously. They often face difficulties in tracking and organizing tasks, making sure everyone is aware of their respective responsibilities, and ensuring timely communication and collaboration across teams and departments. \\
\hline
0.6 &   The company provides a comprehensive project management solution that helps organizations manage their projects effectively. By facilitating collaboration, tracking progress, and providing visibility into the overall project status, they enable teams to work together efficiently, stay on track, and ensure timely completion. Their solution also offers features such as task management, time tracking, project milestones, and reporting tools, making it easy for teams to track their progress, monitor deadlines, and make necessary adjustments. Furthermore, their platform integrates with other tools and systems, further streamlining communication and ensuring alignment across teams. Overall, this solution helps organizations optimize project management, improve productivity, and achieve desired outcomes. & Many organizations struggle with managing their project management and collaboration efforts effectively. They often face challenges in coordinating teams, tracking progress, and ensuring timely project completion. \\
\hline
0.7 &   A company offers a comprehensive cloud-based collaboration and project management platform that helps organizations streamline communication, enhance team productivity, and overcome disparate collaboration challenges. Their platform provides a centralized hub for organizing and organizing team communication, ensuring seamless collaboration across devices and platforms. It offers features such as online file sharing, project management tools, task management systems, and real-time chat channels, all integrated into a unified platform. This solution not only improves collaboration and productivity but also ensures that projects are executed effectively and efficiently. By offering a seamless, user-friendly interface, this company empowers teams to work together efficiently and effectively. & Many companies struggle with ineffective collaboration and productivity issues. They often face challenges in organizing and maintaining communication across different teams and platforms, resulting in disjointed workflows and a lack of efficiency. \\
\hline
0.8 &   The company provides a solution that addresses these challenges by leveraging cloud-based tools and communication platforms. By providing a centralized platform, companies can empower their teams with tools for organizing project tasks, setting priorities, tracking progress, and communicating updates effectively. This centralized system eliminates the need for multiple tools or systems, reducing the risk of data loss or miscommunication, and enabling smooth project coordination and overall success. Additionally, the software facilitates real-time collaboration with team members, allowing for seamless communication across different locations and time zones. Overall, the solution helps companies optimize project management, improving productivity and achieving success in their overall goal. & Many companies struggle with organizing and managing their project-based work and communication efficiently. They face challenges such as difficulties in tracking tasks, managing priorities, and ensuring timely updates across teams or locations. \\
\hline
0.9 &   The company offers a comprehensive platform that facilitates team collaboration and project management. Their software enables streamlined communication, real-time file sharing, and automatic project tracking. With this solution, organizations can improve productivity and efficiency, allowing teams to focus on their mission without having to worry about disjointed tasks. Additionally, the platform provides constantly updated data and insights, ensuring transparency and understanding of tasks progress. This solution empowers and supports efficient and responsive project management, leading to better outcomes for organizations and their stakeholders. & Many companies and organizations struggle with inefficient communication, disjointed task management, and lack of real-time collaboration. This creates difficulties in team coordination, project management, and accurate tracking of tasks. \\
\hline
1.0 &   The company's solution offers a comprehensive software solution designed to streamline and automate the product development cycle. It includes tools for engineering design and metrology analysis, as well as collaboration tools for collaboration across teams and inside organizations. Through these tools, companies can effectively manage engineering tasks, such as toolpath planning, nesting, six-sigma validation, surface modeling, fabrication, and pre-production inspection. Additionally, the solution automates the deployment of resources by allowing for faster iteration cycles and precise planning, minimizing the risk of costly errors and optimizemobility. With its robust solution, companies can improve productivity, enhance innovation, and deliver better products faster. & Companies involved in the production of physical goods face numerous challenges in delivering high-quality products on-time and within budget. One of the key obstacles is the lack of efficient and streamlined methods for delivering product designs into reality. \\
\hline
1.1 &   Our company offers a comprehensive software development platform that streamlines the whole process. Our platform provides scalable agile development tools and features, such as project management and Bug Tracking Systems. By integrating existing systems and pull requests, it enables team members to collaborate effectively and efficiently create comprehensive and verifiable requirements.With our solution, companies can easily capture their requirements, identify common concerns, prioritize features, and define timelines. Through continuous deployment, we automate the production and rollout of new features in a continuous delivery cycle. Our platform integrates with various software infrastructure tools to facilitate collaboration with internal and external resources. & companies often struggle with efficiently capturing and incorporating their software development requirements in their processes and systems. This often leads to a disjointed approach and a lack of clarity on priorities. \\
\hline

\end{tabular}
\end{center}
\end{table}

\newpage
\section{Problem Statements in Exploration Experiments}\label{problemstatements}

\begin{table}[htbp]
  \caption{Original problem statements used in exploration experiments.}
\begin{center}
\begin{tabular}{|p{.95\linewidth}|}
\hline
Software project timelines are often underestimated, which leads to high costs. \\
\hline
It is difficult to measure employee satisfaction in an unbiased way. \\
\hline
It is not easy for early startups to find a customer base willing to try new technology. \\
\hline
Companies struggle with gaining insights from large volumes and high velocity of data. \\
\hline
It is hard to track and measure customer satisfaction across large geographies. \\
\hline
It is difficult to plan investments in an uncertain economy. \\
\hline
It is difficult to create innovation opportunities without introducing too much process and hampering creativity. \\
\hline
Retaining high-performing talent is hard in competitive emerging markets. \\
\hline
Large machine learning models are expensive and time consuming to train. \\
\hline
Ensuring privacy of customers is difficult while leveraging their data for business insights. \\
\hline
\end{tabular}
\label{T:origprob}
\end{center}
\end{table}

\newpage
\section{Exploration and Temperature}\label{exploration}
Here we show how our exploration is impacted by LLM temperature.
The exploration uses the following procedure to obtain 100 novel generated solutions starting from a single prompt.

\begin{figure}[htp]
        \centering
                \includegraphics[width=\textwidth]{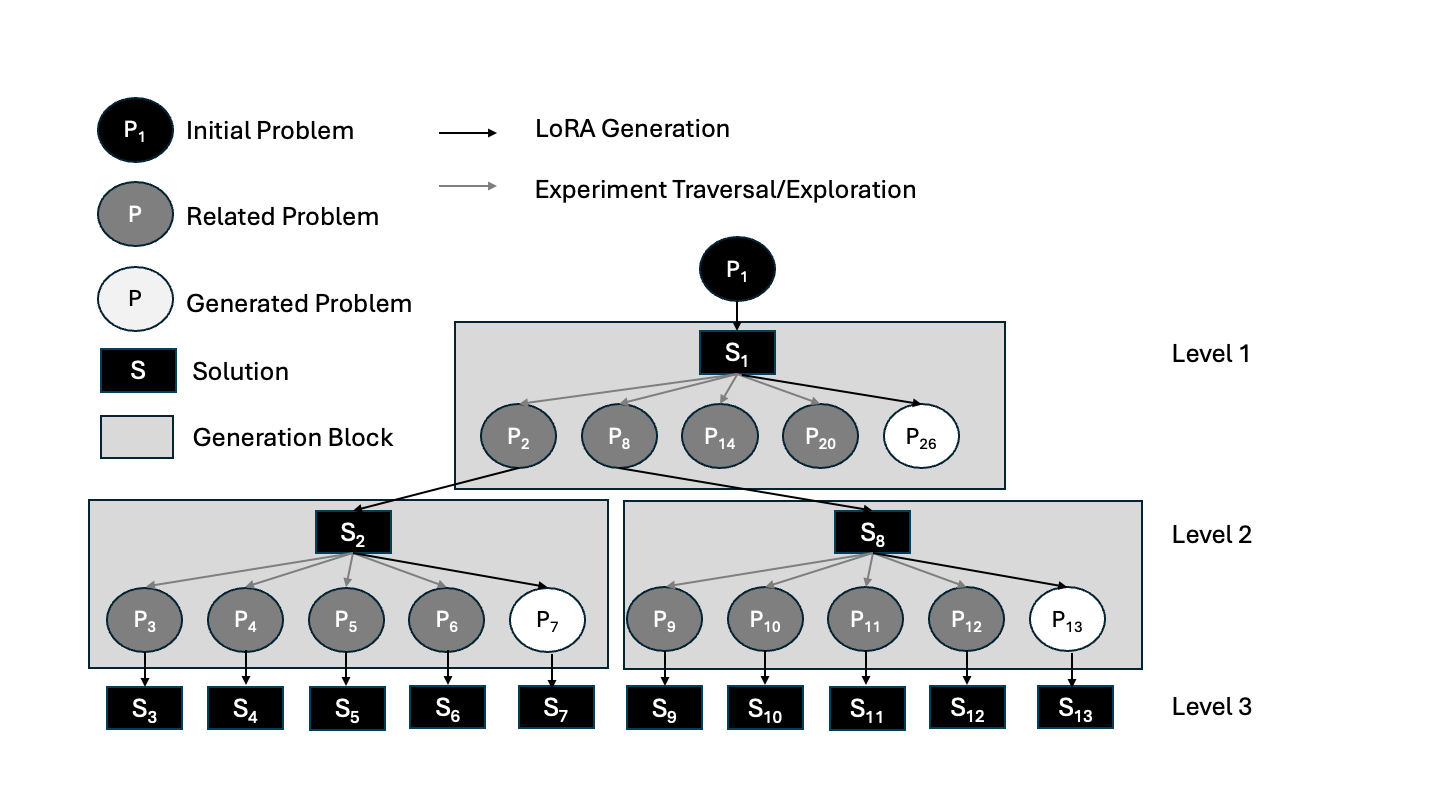}
	\caption{3-level depth-first traversal of problem-solution tree}
	\label{depthtree}
\end{figure}

\begin{enumerate}
\item{Feed the problem into the prompt of the {\it ProblemSolution} LLM adapter and ask for a solution akin to the format used during LoRA tuning.}
\item{Do a nearest neighbor search of the top 4 related problems in our vectorstore}
\item{Feed the solution generated in step 1 as a prompt into the reverse {\it SolutionProblem} LLM adapter to obtain a novel problem}
\item{Recursively depth-first explore all five problems re-starting with Step 1}
\item{When 100 solutions are generated exit the recursion and compute statistics over the solutions and problems obtained}
\end{enumerate}
Two statistics are computed in the final step for both the list of problems and solutions obtained. 
The average edit distance (Levenshtein) between each unique pairing of solution-solution (or problem-problem)
and the average cosine similarity for the embeddings of each unique pairing of solution-solution (or problem-problem).
Max depth was set to 6. A depth-first search with 3 levels is illustrated in Figure~\ref{depthtree}.

Now we repeat this evaluation while modifying the LLM temperature. 
Examples of solutions generated from problems, and problems generated from solutions with this approach at different temperature levels for the
dataset are shown in Appendix~\ref{solution} and Appendix~\ref{problem} respectively.

We computed the edit distance and cosine similarity metrics with increasing temperature for
solutions as well as problems for 10 different original problem statements in 10 separate explorations (traversal trees). 
Edit distance 0 and cosine similarity of 1 mean that the two solutions
compared are the same from a text/word perspective, and semantic/embedding perspective respectively. So for our purposes we want the average
edit distance to be as high as possible and the cosine similarity to be as low as possible to obtain more novelty in the solutions.
At the same time we do not want the solutions to drift too far away semantically.

\begin{figure}[htp]
        \centering
        \includegraphics[width=2in]{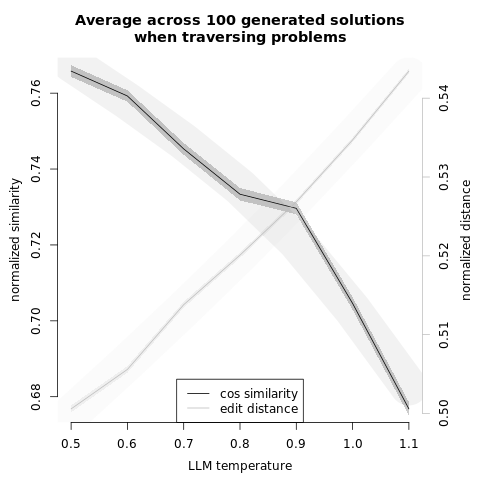}
        \includegraphics[width=2in]{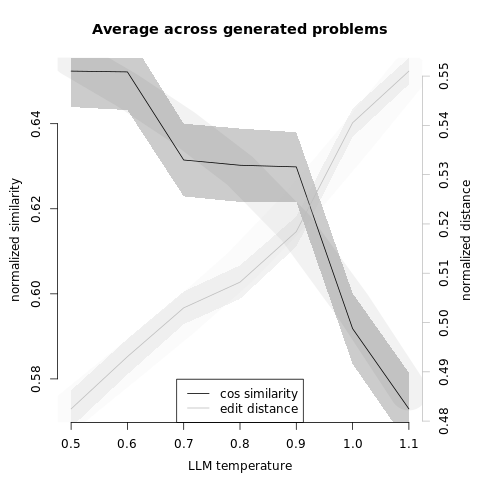}
	\caption{Solution and Problem Novelty: Average across all original problem statements. See Appendix~\ref{problemstatements}.}
	\label{compsolcombined}
\end{figure}

We can see in Figure~\ref{compsolcombined}
that solution novelty, defined here lexically as the edit distance\footnote{increasing with temperature} and 
semantically as the inverse of cosine similarity\footnote{decreasing with temperature}, increases
as expected with temperature. Each exploration at each temperature level involves 100 solution generations, 100 problem generations and 100 nearest neighbor
searches, so this Figure is based on data from a total of $14,000$ LLM generations, and $7,000$ nearest neighbor searches.

We also observe that for the problem space the same technique works and we get the same novelty
effect on generated problems when increasing the temperature.

\newpage
\section{Pertinent Survey Questions}\label{surveyquestions}
\subsection*{Phase 1 Survey Questions}\label{phase1questions}
\begin{table}[htbp]
  \caption{Phase 1 survey questions asked of users after an initial 15 min onboarding process that involved a demo of the tool and guiding the user through a structured use of the tool.}
\begin{center}
    \begin{tabular}{|p{.75\linewidth}|p{0.20\linewidth}|}
        \hline
        \textbf{Survey Question} & \textbf{Answer Type} \\
\hline
How would you rate your use of GenAI tools like ChatGPT, CoPilot, or Gemini? & Quant\\
\hline
How likely are you to use AI-deation for a future brainstorm with yourself?  & Quant \\
\hline
Why did you rate AI-deation the way you did above? Please include any discussion of where we can improve the tool or examples where is consistently surprised or disappointed you.  & Open Ended\\
\hline
\end{tabular}
\label{T:survey1}
\end{center}
\end{table}
\subsection*{Phase 2 Survey Questions}\label{phase2questions}
\begin{table}[htbp]
  \caption{Phase 2 Survey questions asked after users integrated the tool into their innovation workflows for over a week. Users could explore problem statements of their own choice, and at their own pace.}
\begin{center}
\begin{tabular}{|p{.75\linewidth}|p{0.20\linewidth}|}
    \hline
    \textbf{Survey Question} & \textbf{Answer Type} \\
\hline
How would you rate the quality of solutions proposed by AI-deation? & Quant\\
\hline
How likely are you to use AI-deation for a future brainstorm with yourself? & Quant\\
\hline
Why did you rate AI-deation the way you did above? Please include any discussion of where we can improve the tool or examples where is consistently surprised or disappointed you. & Open Ended\\
\hline
There are multiple ways to use the tool after submitting the initial problem statement.Please rank the usefulness of the following strategies in terms of traversing and exploring a problem deeper. & Multipart \\
\hline
\item[$\rightarrow$] Generating multiple new solutions to the same original problem. & Quant\\
\hline
\item[$\rightarrow$] Generating a new problem based on the initial solution to explore further. & Quant\\
\hline
\item[$\rightarrow$] Based on initial solution, selecting an existing related problem to explore further. & Quant\\
\hline
\item[$\rightarrow$] Submitting a new, modified version of the original problem to explore. & Quant\\
\hline
How easy and valuable was it to integrate this tool into your current ideation process? & Quant\\
\hline
How do you ideate today? How does this tool fit into your current process?  & Open Ended\\
\hline
How would you rate the usability and UI of the AI-deation tool? & Quant\\
\hline
Overall feedback and suggestions for the UI, especially in terms of functionality. What did you find intuitive or confusing about the UI? What improvements would you suggest?  & Open Ended\\
\hline
Any other feedback to the AI-deation team?  & Open Ended\\
\hline
\end{tabular}
\label{T:survey2}
\end{center}
\end{table}


\end{document}